\documentclass[reprint,amsmath,amssymb,prb]{revtex4-1}

\usepackage{bbding}
\usepackage[utf8]{inputenc}
\usepackage[T1]{fontenc}
\usepackage{graphicx}% Include figure files
\usepackage{dcolumn}% Align table columns on decimal point
\usepackage{bm}% bold math
\usepackage{multirow}%Para las tablas
\usepackage{subfigure}
\usepackage{hyperref}% add hypertext capabilities
\usepackage[american]{circuitikz}
\usepackage{tikz}

\DeclareMathAlphabet{\mathpzc}{OT1}{pzc}{m}{it}

\begin{document}

\title{Stochastic Resonance in Neural Network, Noise Color Effects.}
\author{Alexandra Pinto Castellanos}
\address{Zurich, Switzerland}
\email{apintoca@student.ethz.ch}

\date{September, 2011}% It is always \today, today,
\begin{abstract}
Some systems cannot be predicted by classical theories and require  the development of combined deterministic and stochastic theories that make used of noise for their dynamical prediction. Noise is not always an interfering signal which perturbs the system. On the contrary, noise signal can enhance the performance of non-linear systems. The advantage of noise can be observed through Stochastic Resonance (SR), where noise is used for amplification and subsequent detection of small signals. To detect this phenomena it is necessary that the systems have a bistable potential barrier that creates a threshold, the input of the system should be a weak periodic signal which's amplitude is below threshold together with an stochastic signal. This way, SR is detected when there are weak periodic signals that are added to different noise colors, in order to be amplify and optimise. The interaction between the two signals transforms the potential of the system precisely to the frequency of the weak periodic signal which is added to the system. The behaviour of the SR is detected in a neural network and it is studied under noise color variations. Here its found that Pink noise amplifies the sub-threshold input signal twenty times more in comparison to white noise. This is evidence of the functionality of background noise in the brain, where neurons are  naturally embedded in pink noise.
\end{abstract}

\maketitle

\section{Introduction}
In theoretical and experimental physics, are common practice to do linear approximations to solve non-linear problems. These approximations have given satisfactory results, given that until certain margin of error, theoretical analysis and linear numerical methods agree with observed experimental findings\cite{Mantegna2000}. However, measurements are subject to the same linear approximations as numerical methods which requires the validity of current models. Often, experimental data which does not match with known models, is considered to be subject to noise and eliminated from the average trend of the data. This deletion of the noisy data is done due to the complexity of noise and limited theoretical tools to analyze them. In addition, models and experimental devices are constrained to implicit linearizations.\\ \\
Real physical phenomenas are intrinsically non-linear and their dynamics modulated by a noisy environment. An example is the nervous system, a highly non-linear system  which is embedded in a noisy biological environment dependent on multiple interactions, for example, channel opening which is only triggered precise stimulus like voltage, pH or the binding of a ligand \cite{Zhou2010}. The nervous system is also a detection device whose input signals are strongly immersed in noise, that comes from the external environment. The nervous system is acting as a decoder of noisy input output information by computing small fluctuating noisy conditions. Enhancement and amplification of those fluctuations is a relevant candidate to explain perception \cite{bulsara1996tuning}.
Hence, it is interesting to study Stochastic Resonance (SR), a phenomena property of non-linear systems. SR enabled to have computation with small fluctuating signals. There is also evidence of the role of SR in the functioning of the brain for the detection of weak signals, synchronisation and coherence in neural connections, synapses and behavior in general \cite{gammaitoni1998stochasticLUCA}.\\ \\
For the first time, the SR model was proposed and numerically stimulated by Roberto Benzi\cite{benzi1981mechanism}. Moreover, Luca Gammaitoni studied the SR model and presented the detailed theory behind the stochastic resonance\cite{gammaitoni1998stochasticLUCA}. A special case of SR systems which is called \emph{Ghost Stochastic Resonance} was discovered by Oscar Calvo\cite{calvo2006ghostcircuit}. It occurs for the maximum resonance frequency where the input energy dissapears.
SR is completely different from the resonance observed in linear systems. The most significant difference is the resonance frequency. The resonance frequency is obtained according to the periodic input signal in SR models, while it depends on the structural properties of the system in linear resonance models.\\ \\
There are analytical limitations to the understanding of the dependence of the stochastic resonance, on the parameters of the periodic input perturbation, the noise and the physics of the system. The ammount and dimensionality of these stochastic differential equations and non-linear equations becomes intractable and unsolvable. However, the experimental technique proposed in this research is a powerful tool to better understand and implement SR.\\ \\
SR models can be exploited in many physical systems such as lasers, SQUID and neural networks. In SR models, adding noise to the input stimulus in the sub-threshold regime can enhance the information sensing process in sensory systems. This is a remarkable property of SR models \cite{moss2004SRreviewapplications}.
One of the important applications of SR models based on this property, is for electro-optical devices to enhance their measurement for data acquisition from small signals, which are embedded in noise. An example of this application was done by Bruno Ando \cite{ando1998threshold}. Another application of SR models is analizing them to enhance neural signals. A pioneer in this endeavour is Frank Moss \cite{moss2004SRreviewapplications}, who showed that SR models agree with the neuron models and its properties. He expressed that the sources of noise can be important for coherence in the brain.\\ \\
Non-linear systems are composed of three main components, i.e., a threshold value , a weak input signal to stimulate the system in the sub-threshold regime and a noise signal. These components are always found in nature and their interactions facilitate SR. According to the stochastic differential equations that model the phenomena, it can be concluded that SR depends on the subthreshold periodic input signal, the noise signal and the system structure; however, there is no analytical solution for SR models. Hence, numerical simulations are used to obtain the solution of these models.SR models have an intrinsic error. This error originates from selecting the threshold value and it can cause serious complications during measurements. Hence, this error should be reduced through optimizing SR models \cite{ando1998threshold}.
The following methods are employed in order to optimizing SR models:
\begin{itemize}
\item The Inter-Spike Interval Histogram (ISIH) analysis
\item The Signal to Noise Ratio (SNR) analysis
\end{itemize}
They obtain the optimum points for SR models when the noise value is given. By using these optimum points for the SR model, the input signal is preserved and amplified. On the other hand, different types of noises, i.e., white and color noises can affect non-linear phenomena, differently. Thus, finding type and amplitude of the noise signal which maximizes the performance of the SR model to amplify the input signal is important.\\ \\
According to biological studies, SR plays an important role in the functioning of the brain for detecting weak input signals and synchronization of neural connections \cite{gammaitoni1989stochastic}. It is possible to see the phenomena in neurons, given that its function in the brain is the integration and information processing of electrochemical signals. This constant activity at the interior of the brain creates a background activity, due to the neuronal pulse and the signal transmission that causes fluctuations in the membrane potential creating a real source of noise \cite{reinker2004stochasticphd}. This noise is the responsible for signal amplification by SR.\\ \\ 
The purpose of this paper it to obtain an efficient SR model for neurons in the presence of different types of noises. This model can amplify the weak input signal efficiently for easier detection. Thus, the main contributions of this research are:
\begin{itemize}
\item Investigate the non-linear behavior of neurons can be modeled by SR models. Then SR model for an artificial neuron is employed in order to analyze conveying information of the weak input signal. In spite of using artificial neurons, our long term objective is analyzing biological neurons.
\item Parameter optimization for the neuron model in order to amplify the weak input signal in the presence of different types of noises, i.e., white noise and color noise.
\item Finally, it is shown in SR models, that pink noise is twenty times more optimal than white noise to amplify the weak input signal.
\end{itemize}

The rest of this paper is organized as follows. In Section 2, the theoretical background of the SR model of the required constraints in order to detect the weak input signal are explained. Moreover, two methods for analyzing the output of the SR model are introduced. In Section 3, the detailed methodologies to develop the neural network architecture from electrical devices is presented together with the structure of the pink noise generator. In Section 4, the results of the power spectrum analysis and signal to noise ratio analysis are used to compare the efficiency of different types of noises. Finally, Section 5 contains the concluding remarks.

\section{Theoretical background}

\begin{figure}[t!]
\centering
\includegraphics[width=.5\textwidth,height=.2\textheight]{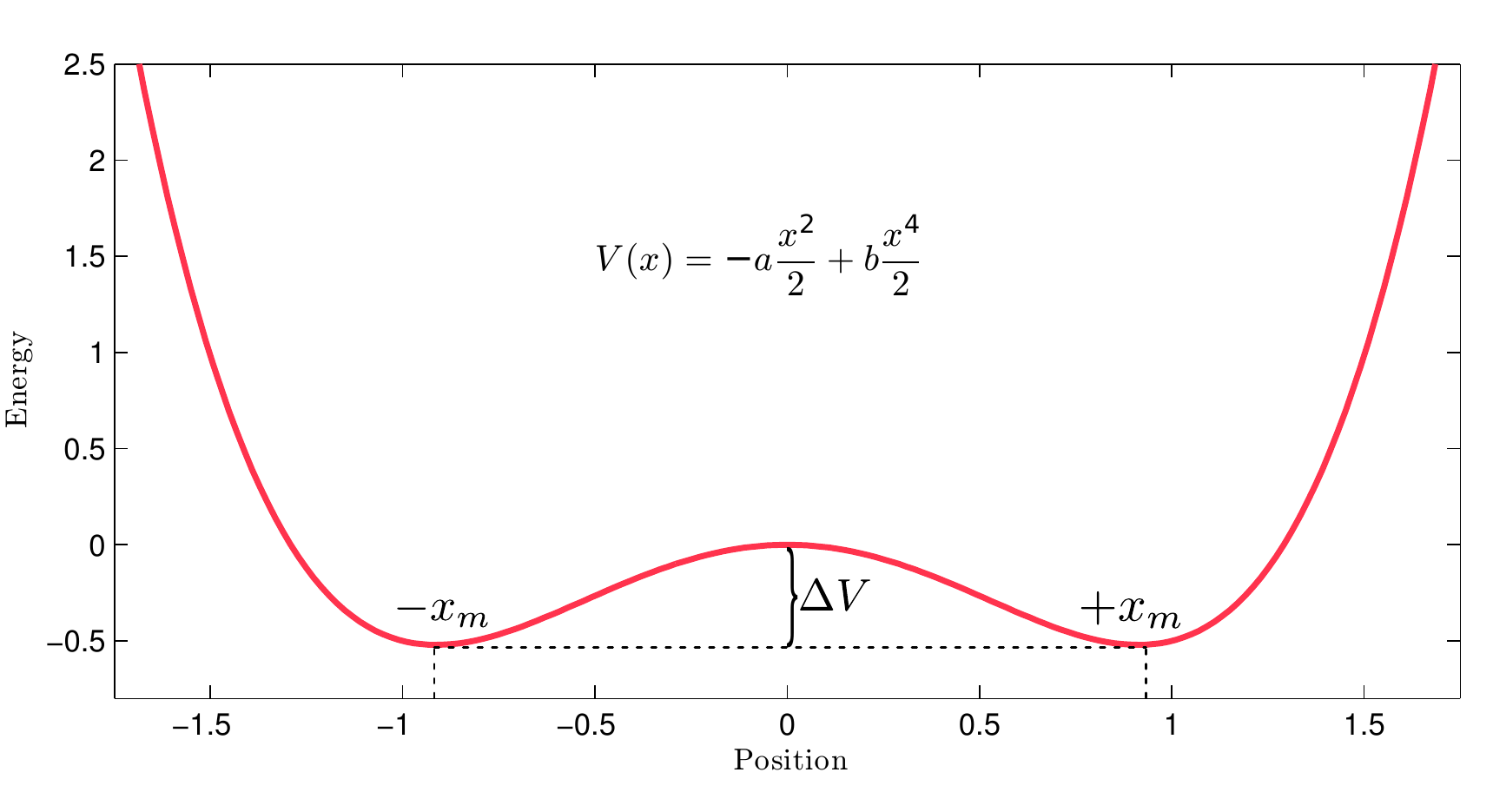}
\caption{Bistable potential which is used in SR model.}
\label{f:Potencial}
\end{figure}

\begin{figure}[t!]
\includegraphics[width=.5\textwidth,height=.4\textheight]{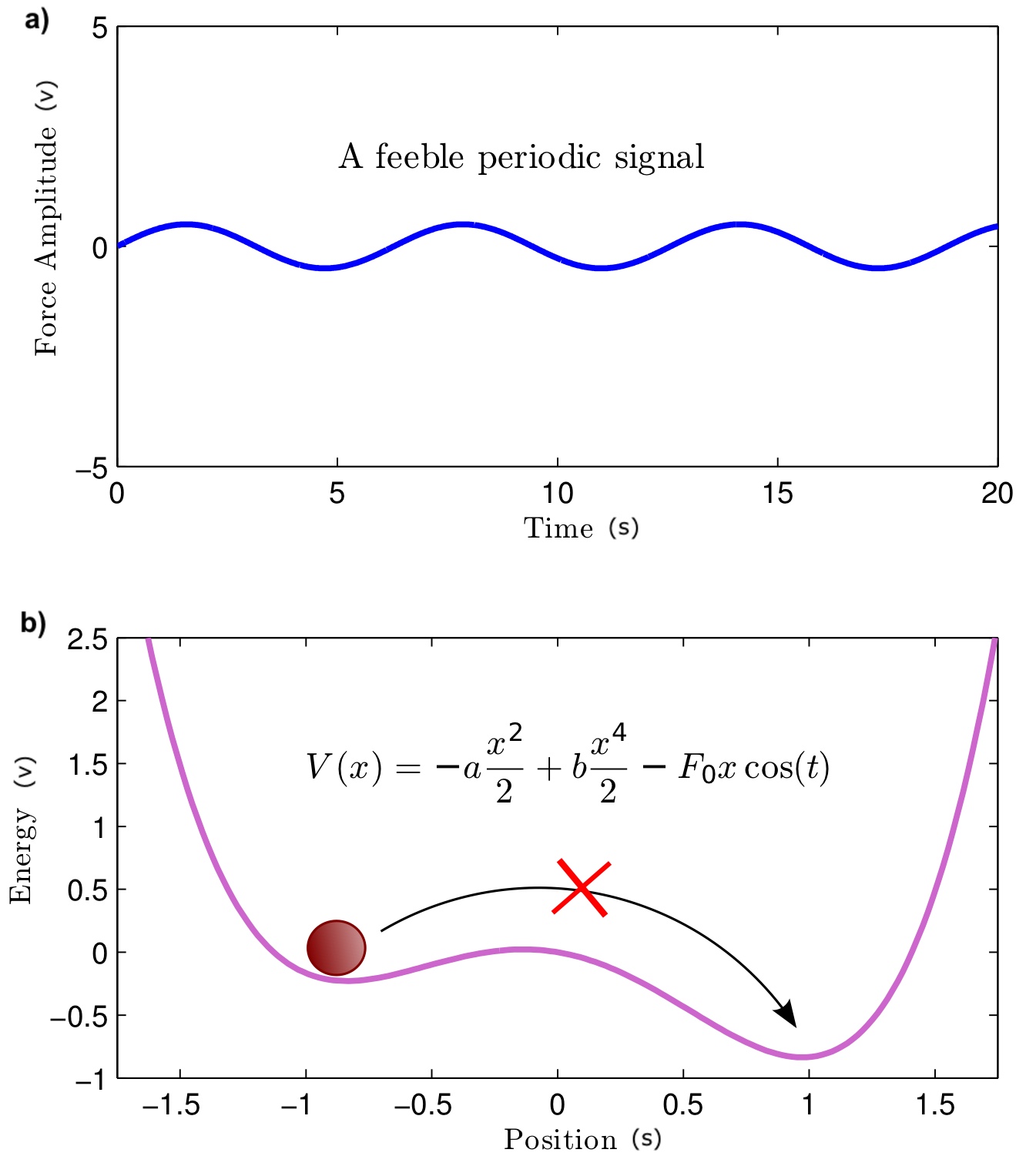}
\caption{a) A weak periodic signal to move the particle. b) The amplitude of this signal is not sufficient to move particle from one side of the bistable potential to the other side.}
\label{f:Sdebil}
\end{figure}

The SR model can be developed through adding a weak periodic force, the variation of a bistable potential function, and a noise signal to the components of the Brownian motion model \cite{Allison2003}. Hence, the SR model can be described as \cite{gammaitoni1998stochasticLUCA}
\begin{equation}
m\frac{d^2x}{dt^2}=F(t)-b\frac{dx}{dt}-\frac{dV(x)}{dx}+\xi(t),
\label{eqn:SRDiffEq}
\end{equation}
where $F(t)$ is the periodic force, $\xi(t)$ is the noise signal and $V(x)$ is the bistable potential which is defined as \cite{ando1998threshold}
\begin{equation}
V(x)=-a\frac{x^2}{2}+b\frac{x^4}{2},
\label{eqn:Pot}
\end{equation}
where $a$ and $b$ are constants and they can modify the shape of potential as shown in Fig. \ref{f:Potencial}.
The constants $a$ and $b$ are related with the potential function characteristics as
\begin{equation}
x_m=\sqrt{\frac{a}{b}}, \,\,\,\,\,\,\,\,\, \Delta V=\frac{a^2}{4b},
\label{eqn:PotPar}
\end{equation}
where $x_m$ and $\Delta V$ are shown in Fig. \ref{f:Potencial}. A weak periodic signal as the input of the SR model is introduced. It is assumed that the amplitude of this signal is not sufficient to move particle from one side of the bistable potential to the other side as shown in Fig. \ref{f:Sdebil}.
\begin{figure}[t!]
\includegraphics[width=.5\textwidth,height=.4\textheight]{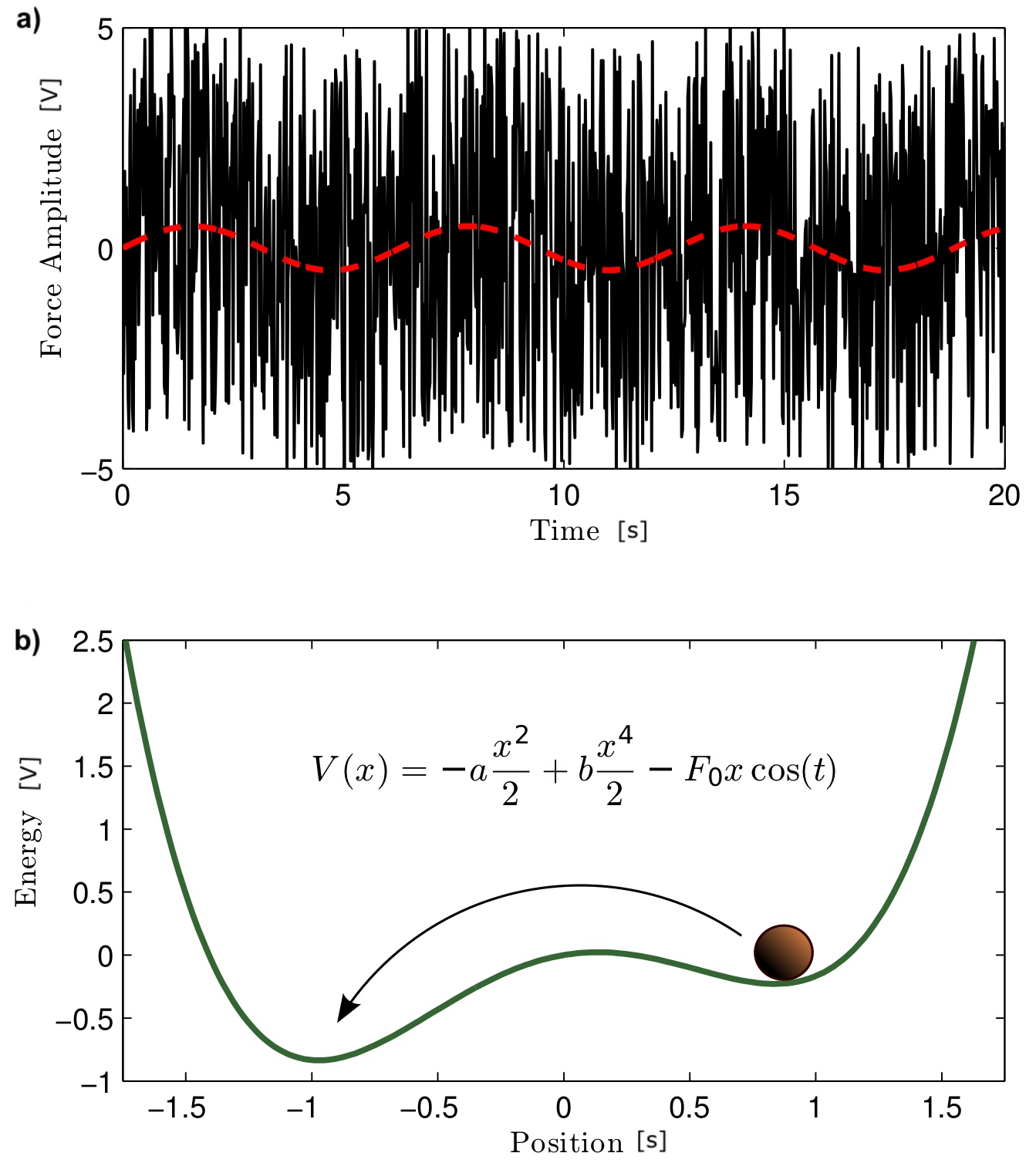}
\caption{a) A noise signal is added to the weak signal. b) The particle can cross from one side of the bistable potential to the other side.}
\label{f:SdebilNs}
\end{figure}
In addition to the weak periodic signal, a noise signal is added to the system. The noise signal enables the particle to cross form one side of the potential to the other side as shown in Fig. \ref{f:SdebilNs}.
\begin{figure}[t!]
\includegraphics[width=.5\textwidth,height=.3\textheight]{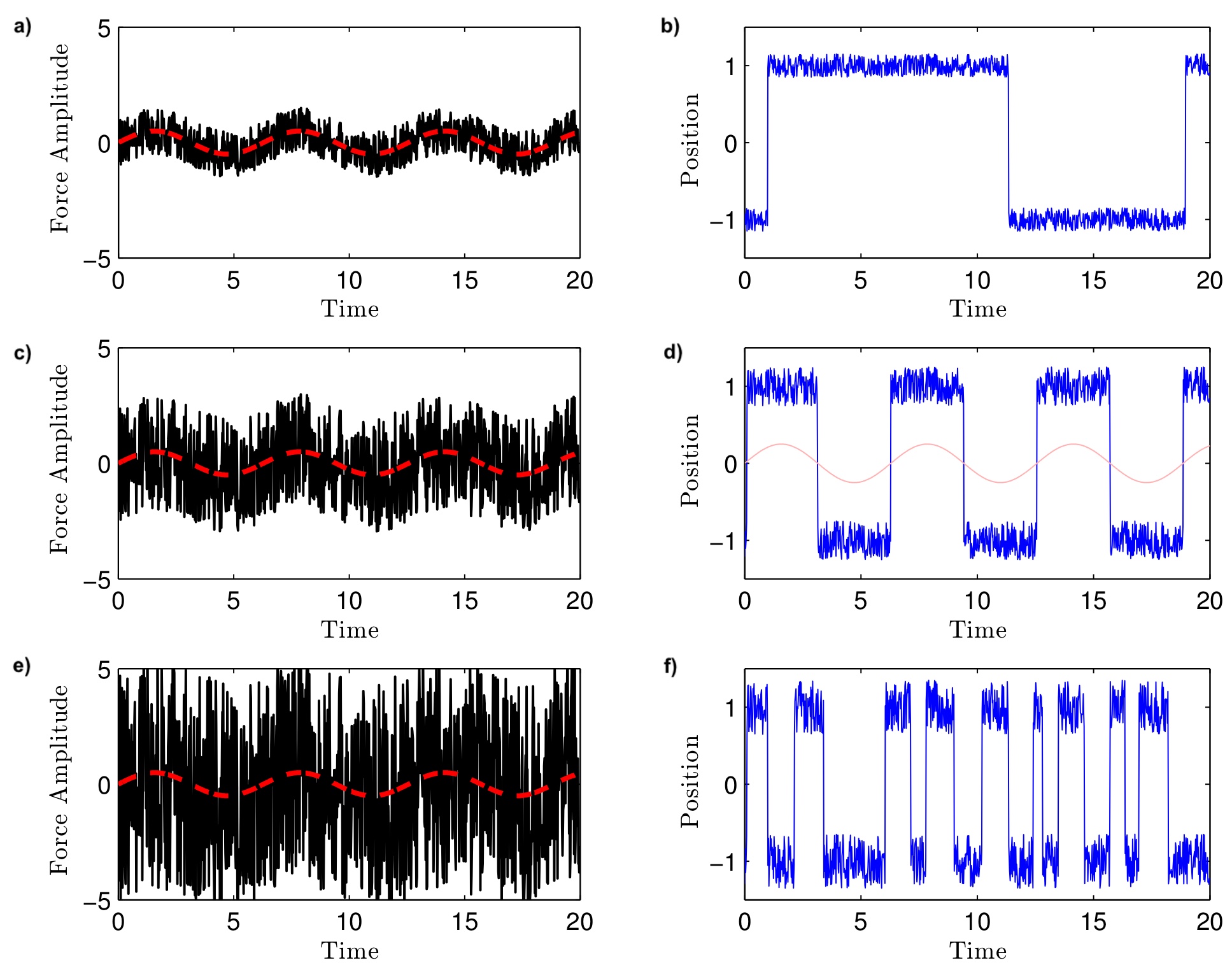}
\caption{Response of the system: (a) Noise amplitude is weak, (b) the number of crosses from one side of the potential to the other side is low. (c) Noise amplitude is optimum, (d) the crosses from one side of the potential to the other side is synchronized with the weak input signal. (e) Noise amplitude is too high, (f) the crosses are so frequent, and thus, the information of the input signal is lost.}
\label{f:Resps}
\end{figure}
Hence, the output of system depends on the noise intensity. When the amplitude of the noise signal is low as Fig. \ref{f:Resps} (a), the rate of moving of article is low as shown in Fig. \ref{f:Resps} (b). On the other hand, when the noise amplitude is high as Fig. \ref{f:Resps} (e), the particle moves form one side to the other side frequently. Thus, in these situations, the input signal is lost. However, when the noise intensity is optimal as in Fig. \ref{f:Resps} (c), the crossing of particle is synchronized with the weak periodic signal as shown in Fig. \ref{f:Resps} (d). The SR model works perfectly in this scenarios.

\begin{figure}[t!]
\includegraphics[width=.5\textwidth,height=.25\textheight]{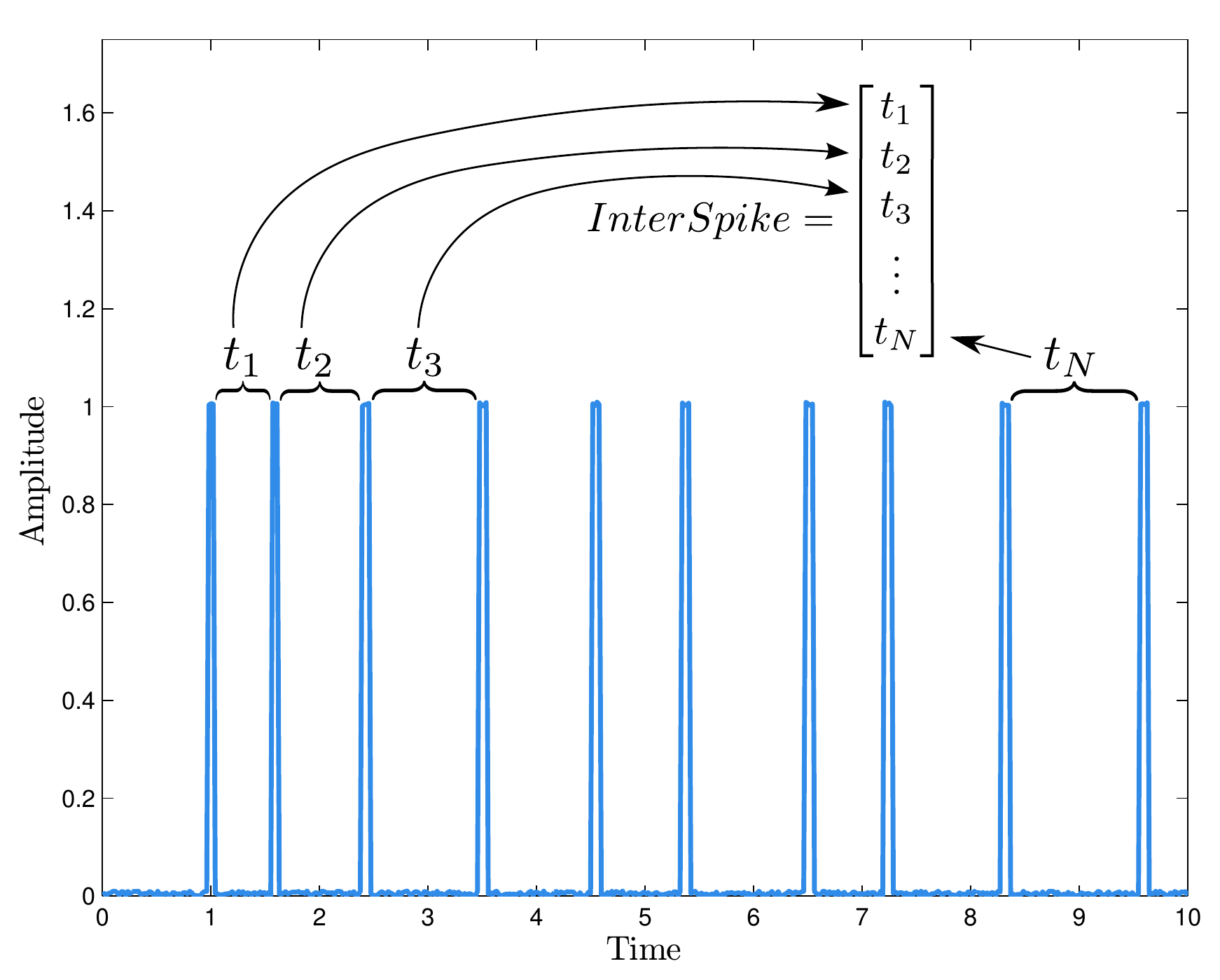}
\caption{ The output of an artificial neuron. The inactivity intervals are measured and saved to analyze.}
\label{f:Spikes}
\end{figure}

Variations of the number of received neuro-transmitters cause fluctuations in the membrane potential and create a source of noise \cite{reinker2004stochasticphd} which can be responsible for signal amplification in the SR model. The output signal of a neuron is a train of Dirac delta functions as shown in Fig. \ref{f:Spikes}.
One way to analyze the SR model in a neuron is measuring the inactivity intervals between Dirac delta functions in the response of a neuron. Then, the collected responses from different neurons are saved for histogram generation \cite{reinker2004stochasticphd}.
\begin{figure}[t!]
\includegraphics[width=.5\textwidth,height=.23\textheight]{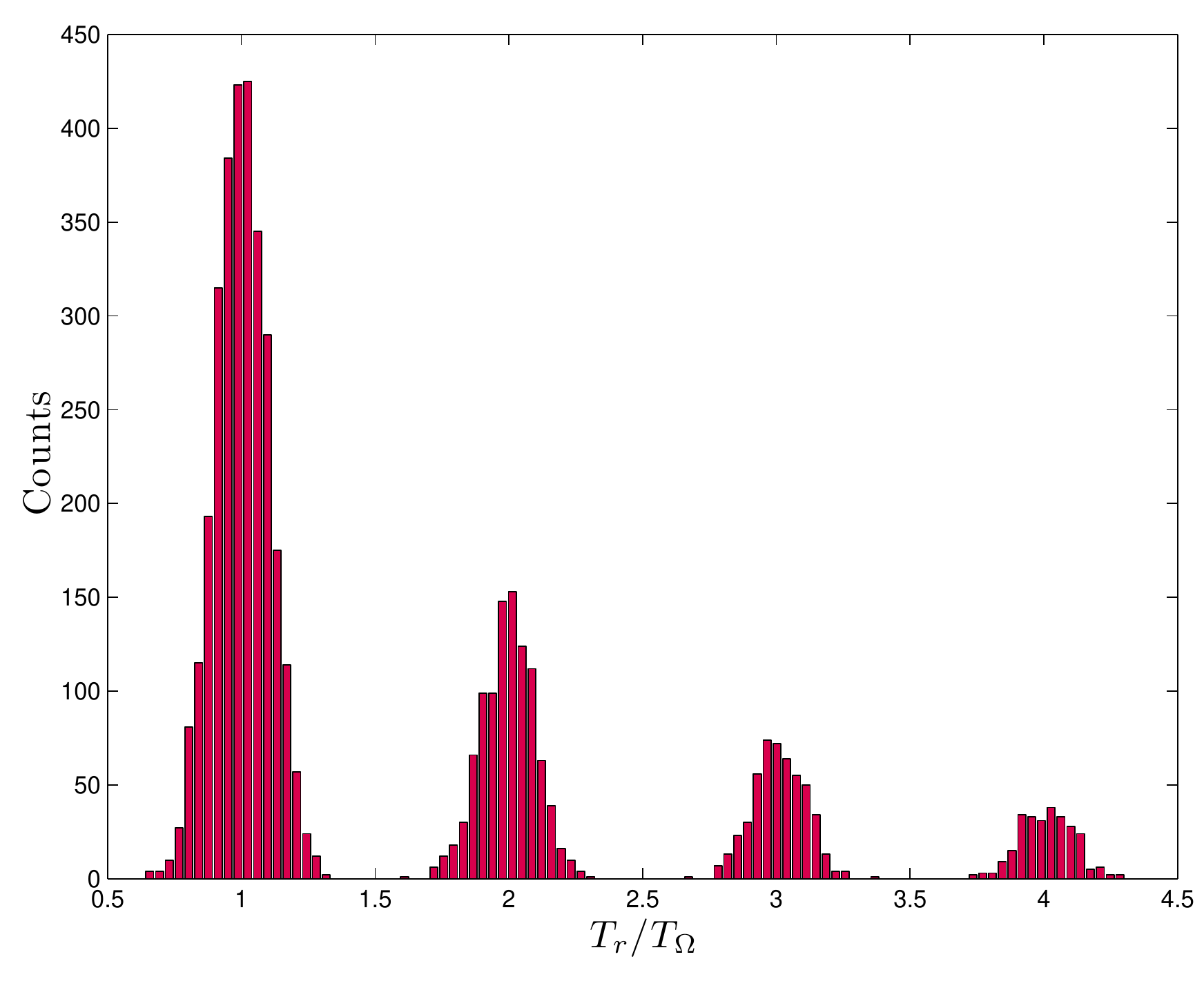}
\caption{Histogram of the inactivity intervals in the output spike train proportional to the input signal's period.}
\label{f:ISIHt}
\end{figure}
Fig. \ref{f:ISIHt} shows the histogram of the time intervals. Variations of the noise signal modify the histogram. When the noise amplitude is low, the inactivity intervals are too long and the majority of intervals are longer than the signal's period. On the other hand, when the intensity of noise signal is high, the inactivity intervals are too short and the majority of intervals are shorter than the signal's period. However, when the amplitude of the noise signal is optimum, the inactivity intervals are equivalent to the times corresponding to the multiples of the input signal period as shown in Fig. \ref{f:ISIHt}. Thus, the SR model is designed based on the maximum value corresponding to the signal's period. The peak of histogram with respect to the noise amplitude is studied, and thus, the optimum intensity of the noise signal for the SR model is obtained. This scheme is called ISIH analysis.\\ \\
Another useful method to quantify SR model is SNR analysis which employs Fourier transform. The Fourier transform of the output signal shows multiple peaks at the multiples of input signal's frequency value. In this method, the peak corresponding to the input frequency is integrated according to the expression in (\ref{eqn:SNRint}) \cite{gammaitoni1998stochasticLUCA} for different noise amplitudes. Then, the derived values is plotted with respect to the noise amplitude to obtain the optimum intensity of the noise signal for the SR model. 
\begin{equation}
\text{SNR}=\left[\displaystyle\lim_{\Delta \omega \to{0}}{ \displaystyle\int_{\Omega-\Delta\omega}^{\Omega+\Delta\omega} S(\omega)\, d\omega}\right].
\label{eqn:SNRint}
\end{equation}

\section{Methodology}
\subsection{Artificial Neuron Circuit}

\begin{figure}[t!]
\centering
\includegraphics[width=.5\textwidth,height=.3\textheight]{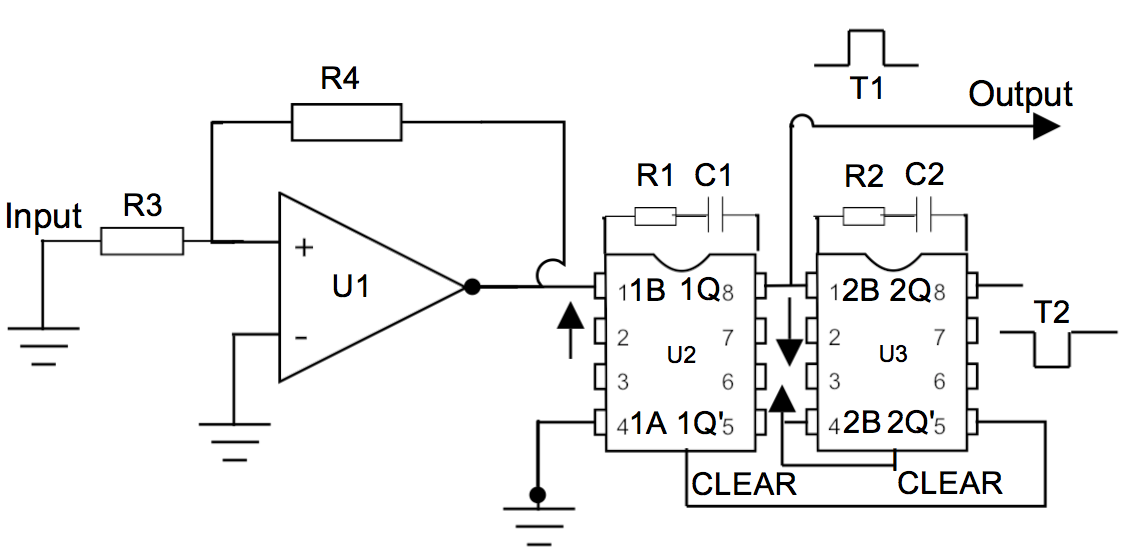}
\caption{Artificial neuron circuit based on the monostable Schmitt Trigger.}
\label{fig:CirDiagram}
\end{figure}

\begin{figure}[t!]
\centering
\includegraphics[width=.45\textwidth,height=.25\textheight]{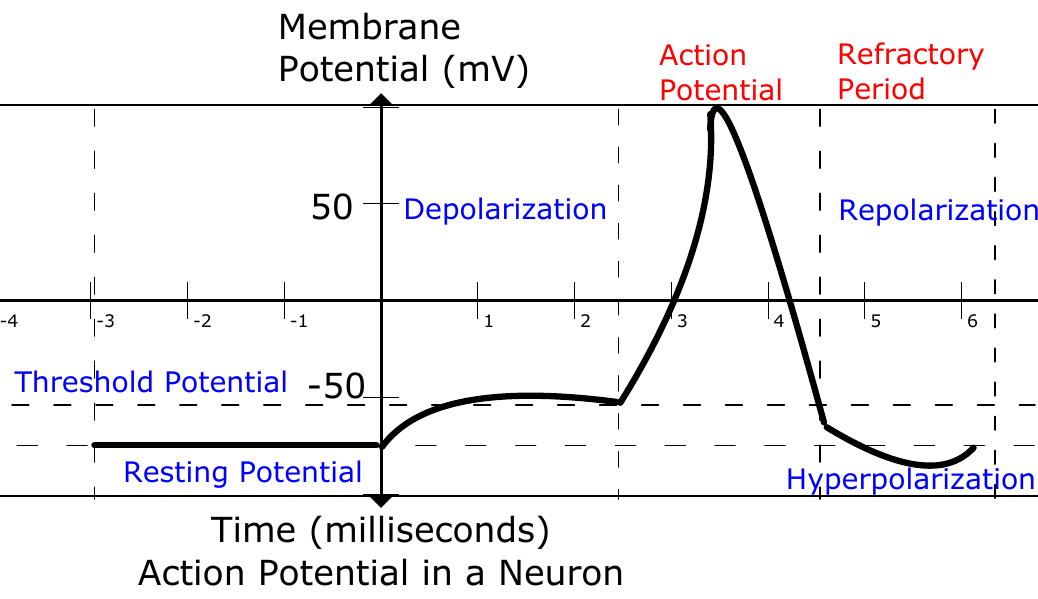}
\caption{The waveform of an action potential. The neuron has a threshold. When the membrane potential overcomes this threshold value, the neuron generates an action potential \cite{crossman2005neuroanatomy}.}
\label{fig:ActionPot}
\end{figure}

\begin{figure}[t!]
\centering
\includegraphics[width=.5\textwidth,height=.5\textheight]{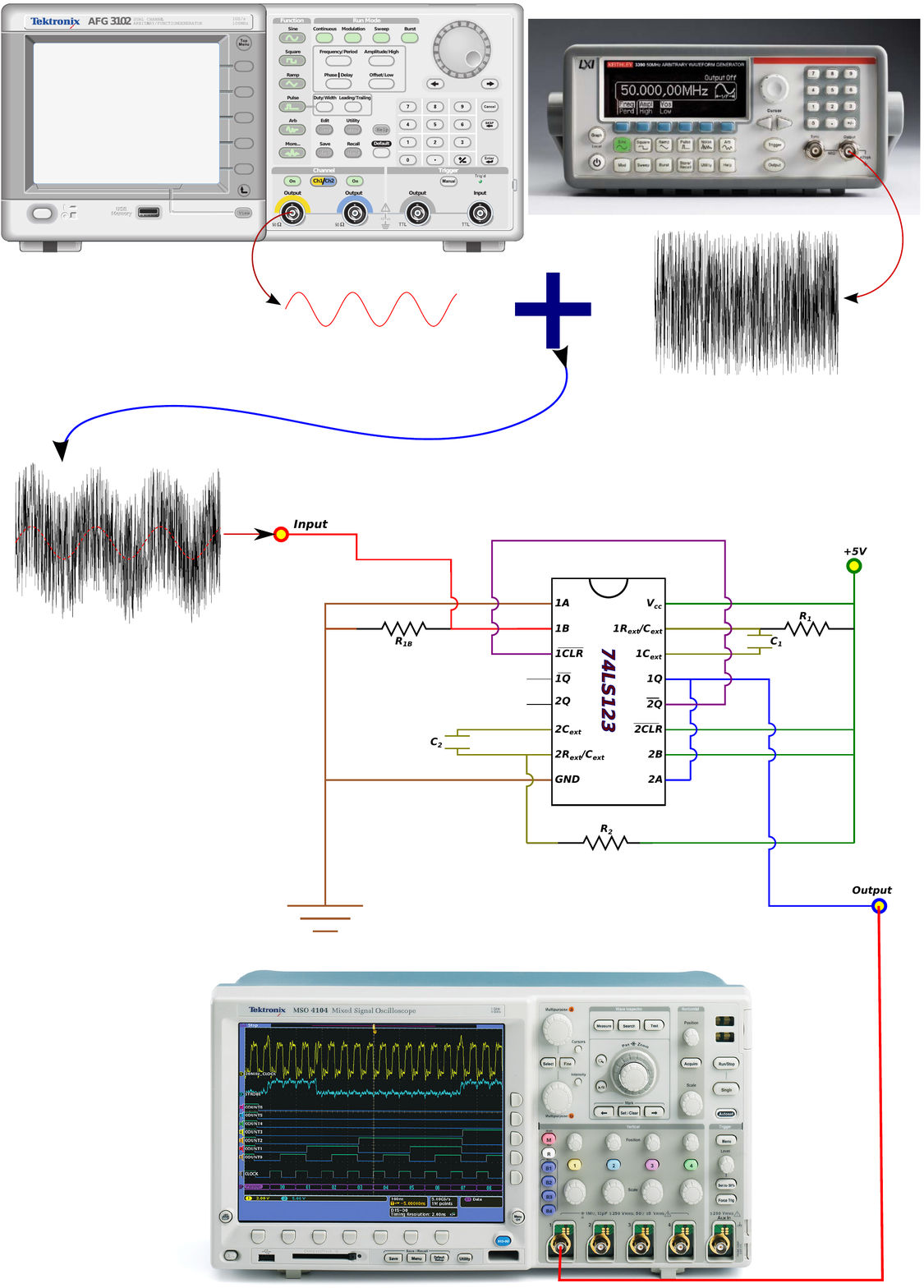}
\caption{Experimental setup for the SR analysis. The upper left device is a coherent signal generator, the upper right one is a white noise generator, the circuit in the center is the artificial neuron and the bottom device is an oscilloscope which is used for data acquisition.}
\label{fig:Setting}
\end{figure}

In this subsection, the employed artificial neuron model for the SR analysis is described. The circuit of an artificial neuron is shown in Fig. \ref{fig:CirDiagram}, as taken from Calvo and D.R. Chialvo \cite{calvo2006ghostcircuit}. It works mainly based on a Schmitt Trigger monostable. This is a device with a non-dynamical threshold value which selects the input signals through comparison with this threshold value. Hence, when the input signal overcomes the threshold value, it generates a spike in the output. \\ \\
The Schmitt Trigger device can simulate the behavior of neurons to convey information through generating action potentials. An action potential is a roughly $100 \,\text{mV}$ oscillation in the electrical potential across the cell membrane which lasts for about $1 \,\text{ms}$, fig. \ref{fig:ActionPot} shows the waveform of an action potential. For a few milliseconds just after an action potential, it may be virtually impossible to generate another spike, this is called the refractory period as shown in Fig. \ref{fig:ActionPot}.
\begin{figure}[t!]
\centering
\includegraphics[width=.5\textwidth,height=.25\textheight]{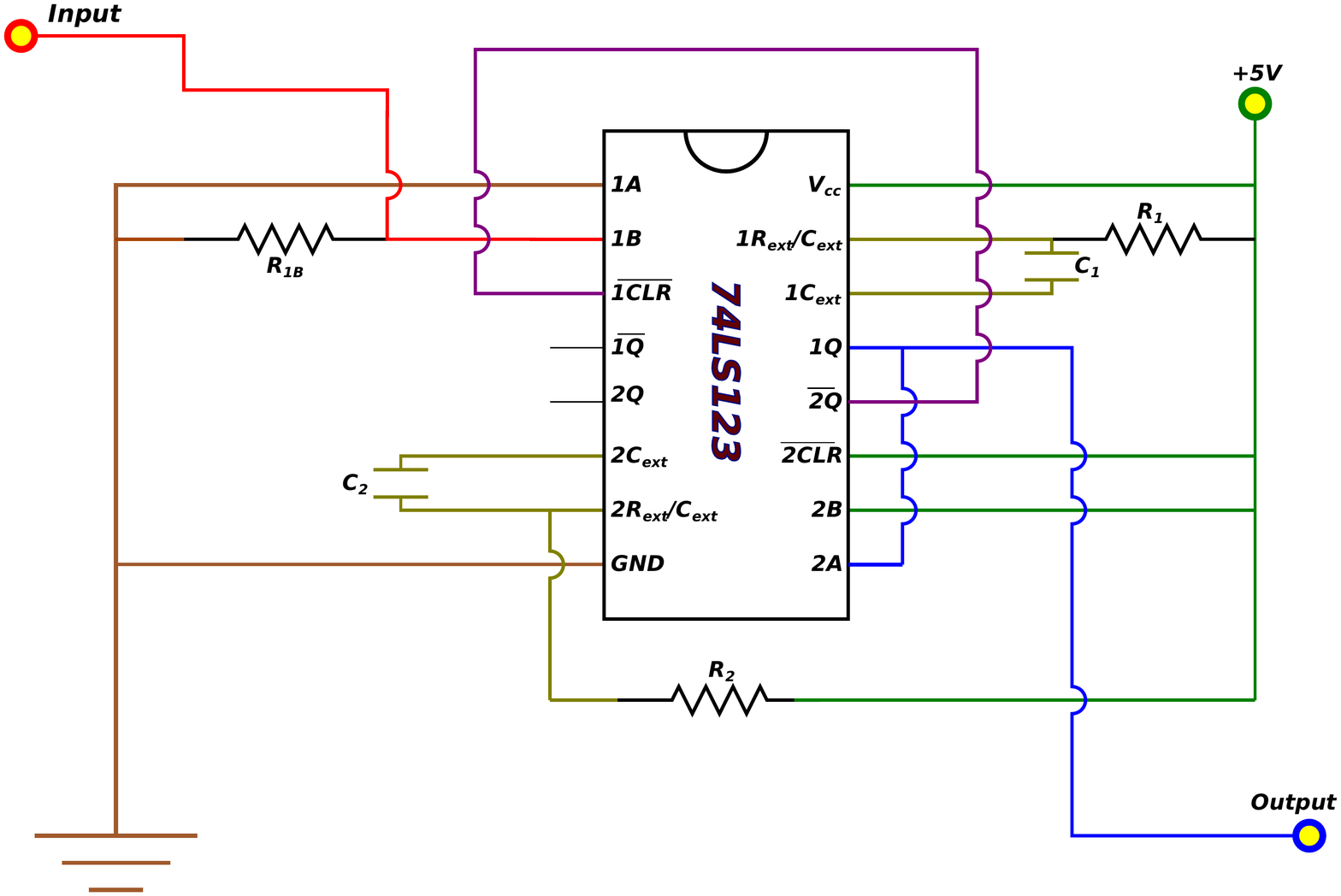}
\caption{Schematic of the assembled artificial neuron circuit.}
\label{fig:Circuit}
\end{figure}
In addition to the Schmitt trigger device, there are operational amplifiers U1 and U2. Unit U1 is an operational amplifier which receives the input signal and amplifies it in order to feed the first monostable through pin 1B. Unit U2 generates a pulse with a duration of T1, when the input signal overcomes its intrinsic threshold. The value of T1 is determined based on the resistance and the capacitance values. The output of unit U2 through pin 1Q simulates the output of the neuron. When the generated pulse by U2 enters in the descendant flank, the second monostable, i.e., unit U3 fires. It generates a pulse with duration T2. The output of pin 2$\bar{Q}$ of U3 works as the clear signal of the first monostable. Hence, U2 does not generate a new action potential in the descendant flank of every spike. This part of the circuit simulates the refractory period of neurons. This cycle is repeated indefinitely until the stimulation signal is present.\\ \\
The experimental setup to analyze the SR model is shown in Fig. \ref{fig:Setting}. A coherent signal generator is used to generate the periodic input signal. This signal is added to the noise signal generated by the noise source. The summation of these signals enters the artificial neuron circuit. This circuit generates an output signal when the input signal overcomes its intrinsic threshold. The output port is connected to an oscilloscope which saves the signal for later processing, in the experimental setup. \\ \\

\subsection{Pink Noise Generation}
\begin{figure}[t!]
\centering
\includegraphics[width=.44\textwidth,height=.32\textheight]{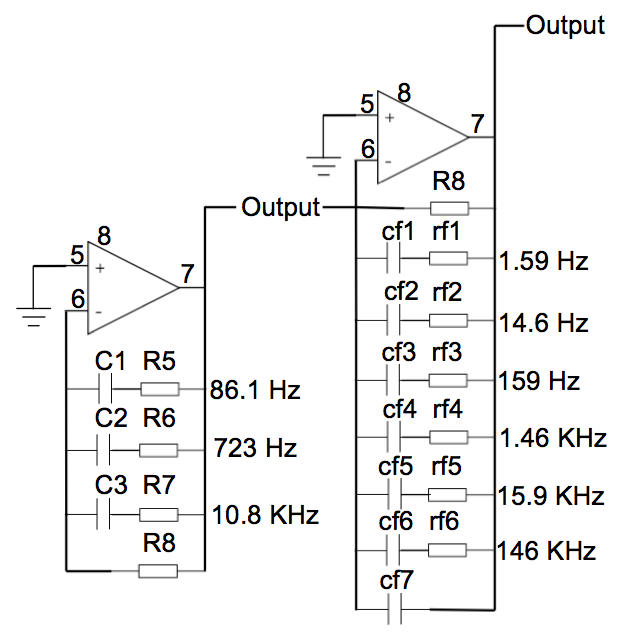}
\caption{Structure of the filter for producing pink noise.}
\label{fig:filtropink}
\end{figure}

In this paper, the color noise is developed by using a pink noise generator. Pink noise is generated through coupling two filters as shown in Fig. \ref{fig:filtropink}. Next, a white noise is used as the input of these filters. Hence, they reduce high frequency components of white noise to produce pink noise. To evaluate the correct acquisition of pink noise, the power spectrum and the characteristic slope in the logarithmic scale is analyzed.
\section{Results and Discussion}
In this section, the results from the SR model analysis are obtained. The obtained results for pink noise and white noise are also compared with each other.
\subsection{Power Spectrum Analysis}
\begin{figure}[t!]
\centering
\includegraphics[width=.5\textwidth,height=.25\textheight]{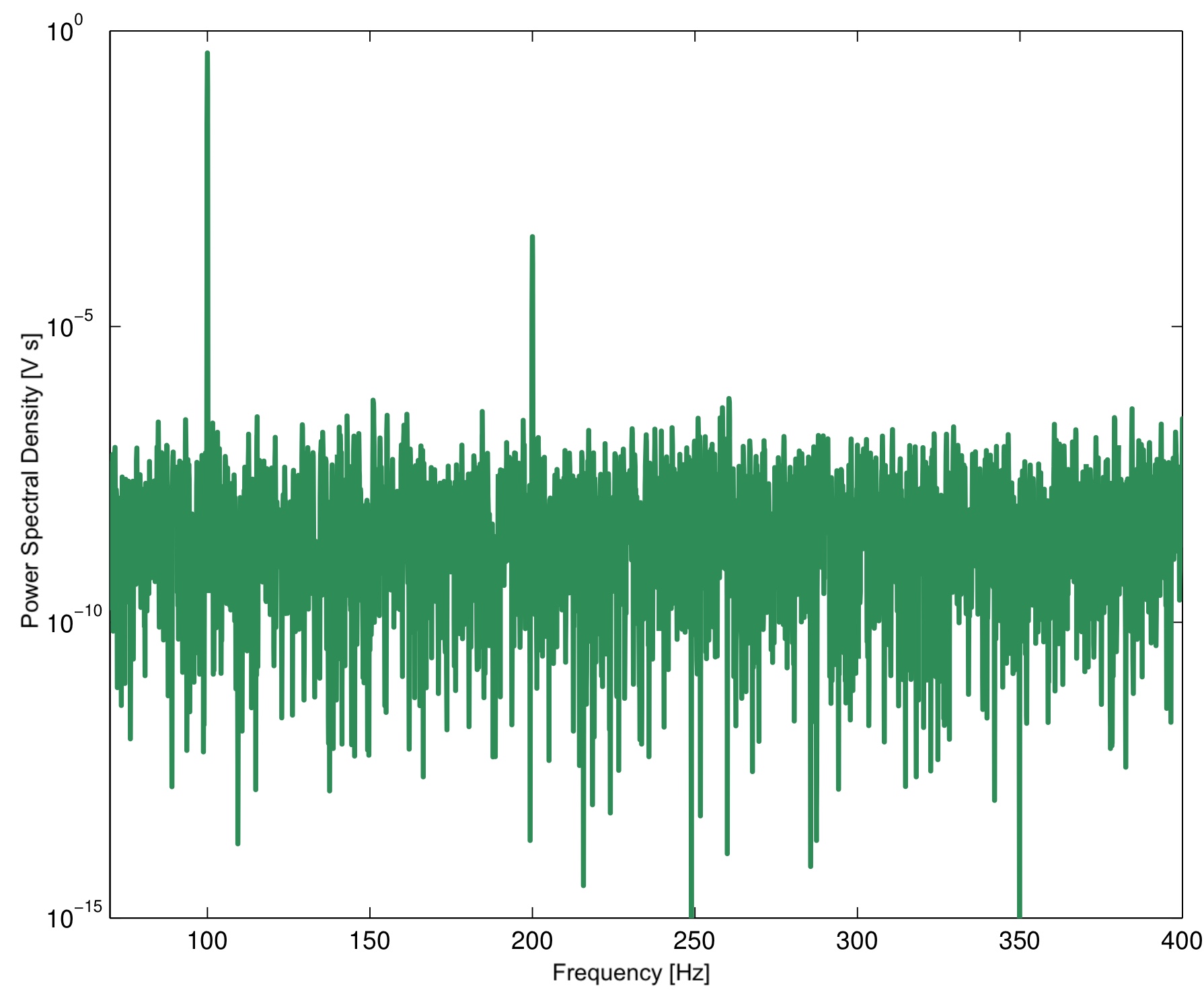}
\caption{Power spectrum of the output signal in the presence of a white noise with $2500 \, \text{mV}$ amplitude.}
\label{fig:FFTW}
\end{figure}

\begin{figure}[t!]
\centering
\includegraphics[width=.5\textwidth,height=.25\textheight]{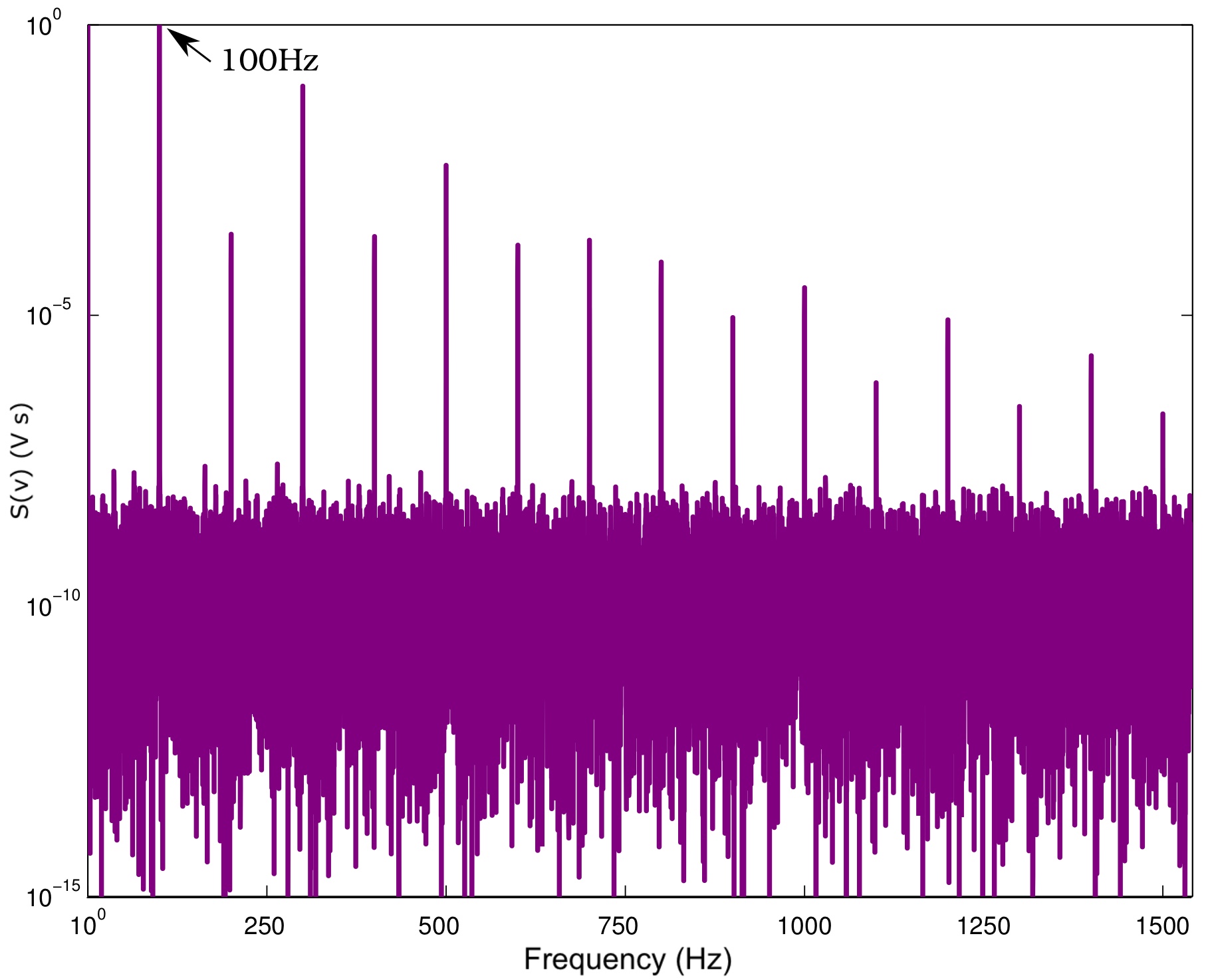}
\caption{Power spectrum of the output signal in the presence of a pink noise with $2900 \,\text{mV}$ amplitude.}
\label{fig:FFTP}
\end{figure}

The Fourier transform is employed for the output signal of the circuit to obtain its power spectrum. It is worth to mention that the captured Fourier transforms by MATLAB are not shown here. Since the numerical results are not as accurate as the obtained results through the spectrum analyzer. Hence, the obtained spectrums by the spectrum analyzer are used for our analysis in the following.\\ \\
Fig.\ref{fig:FFTW} and Fig.\ref{fig:FFTP} show the power spectrums when white and pink noises are used, respectively. In these figures, the X axis corresponds to the frequency and the Y axis corresponds to the power spectrum amplitude. When the peak of the spectrum matches with the inputs signal frequency, it is possible to conclude that the input noise amplitude is optimal. Thus, the noise amplitude is modified to find the optimum values. It can be seen that the number of jumps in the power spectrum is corresponding to pink noise more than in the white noise case.

Through studying different power spectrums, it can be shown that the spectrum amplitude increases when the noise intensity increases. However, when the noise amplitude reaches a specific point, the power spectrum reaches a maximum value. After this point, the peak of power spectrum starts to descend no matter how much the amplitude of pink noise increases.

\subsection{Optimal SNR}

\begin{figure}[t!]
\centering
\includegraphics[width=.5\textwidth,height=.25\textheight]{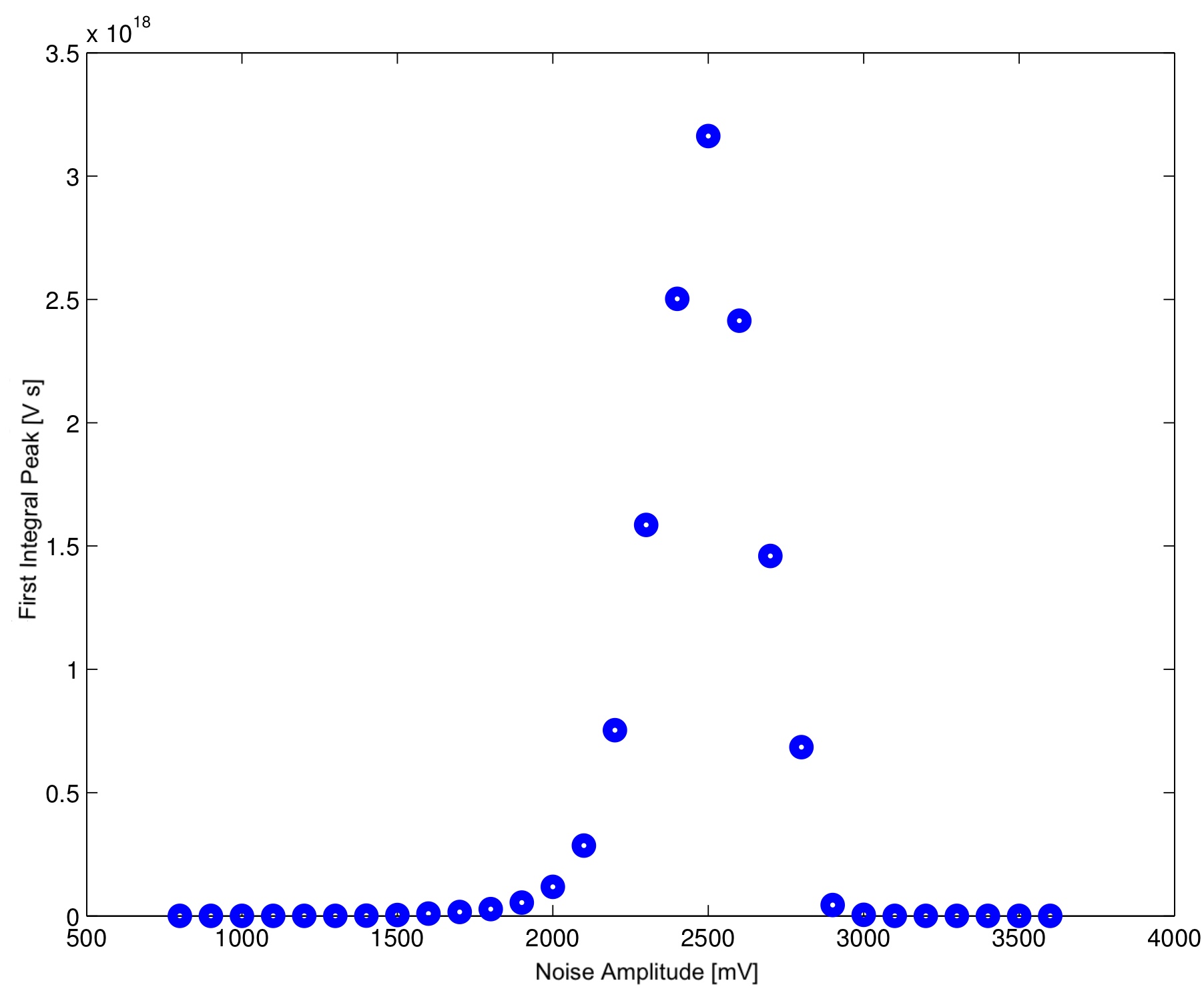}
\caption{Signal to Noise Ratio analysis for white noise when the amplitude of input periodic signal is $50 \,\text{mV}$.}
\label{fig:SNRW}
\end{figure}

\begin{figure}[t!]
\centering
\includegraphics[width=.5\textwidth,height=.25\textheight]{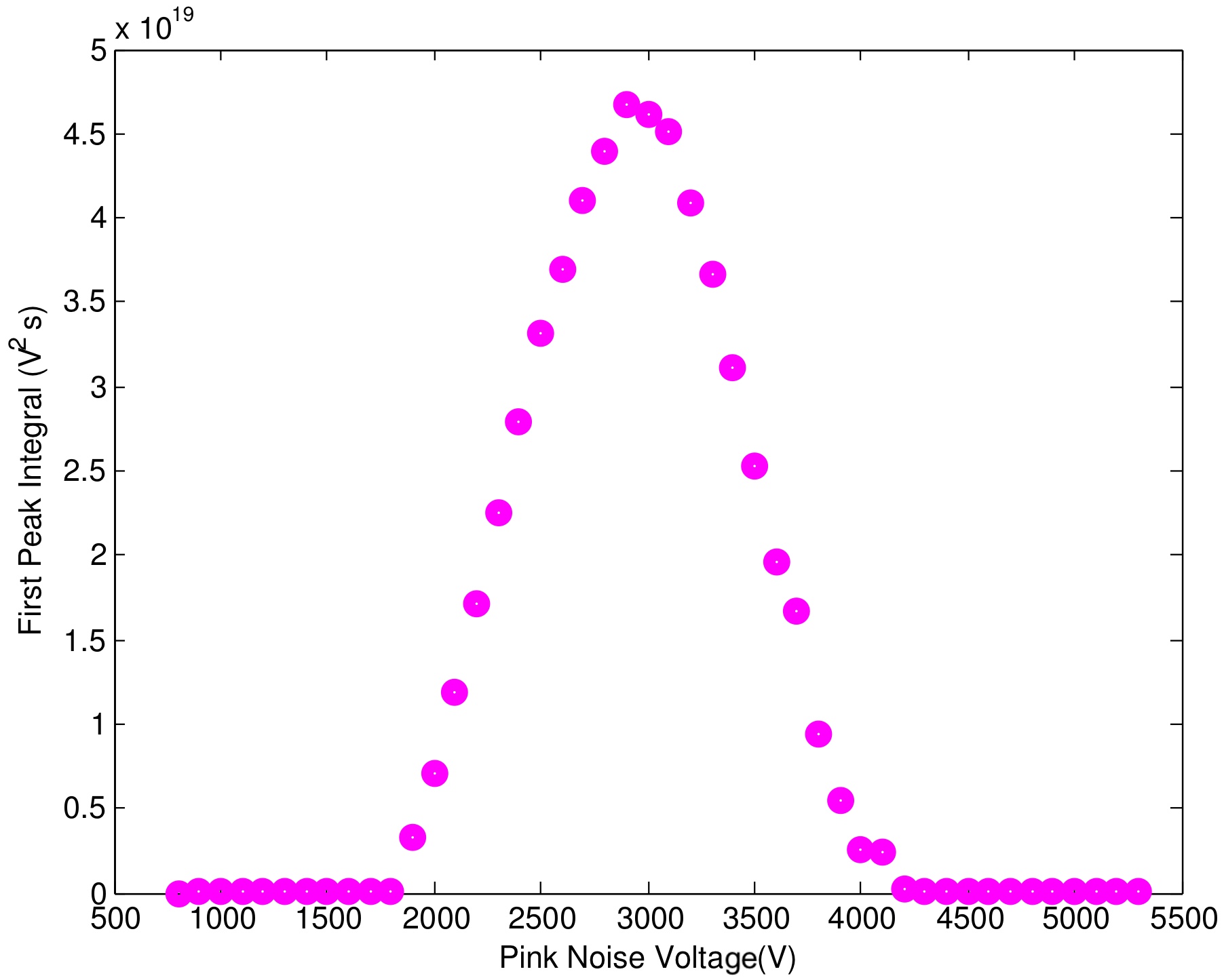}
\caption{Signal to Noise Ratio analysis for pink noise whit an amplitude of  $50 \,\text{mV}$ for the input periodic signal}
\label{fig:SNRP}
\end{figure}

It is necessary to repeat the power spectrum analysis for different values of noise. It can be seen that a maximum peak is conserved. This peak is selected and it is integrated by using the expression in (\ref{eqn:SNRint}) for different values of noise amplitude. The obtained values were plotted with respect to the noise amplitude in Fig. \ref{fig:SNRW} and Fig. \ref{fig:SNRP} for white and pink noises, respectively. These figures show the amplification factor for the input signal versus different values of white and pink noises, respectively.
There is an important difference between white and pink noises. Although the measurements are made for the same voltages and under the same parameters, the maximum performance of SR model is achieved for different optimum noise values. Furthermore, the range of noise values which maximize the SR performance for pink noise is wider than white noise. This range of values for white noise is sharper.

\subsection{Comparison between White and Pink Noises}
\begin{figure}[t!]
\centering
\includegraphics[width=.5\textwidth,height=.5\textheight]{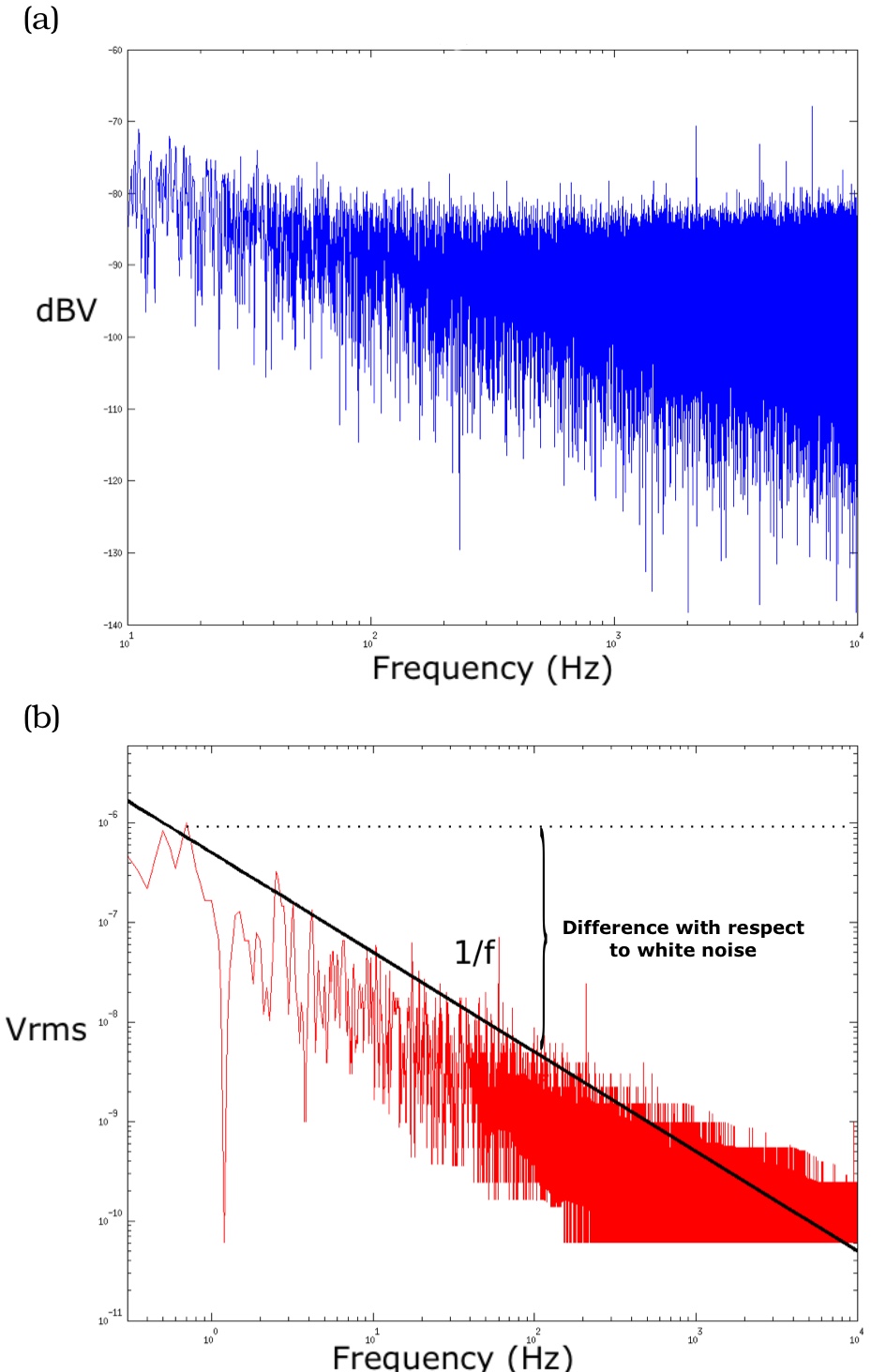}
\caption{(a) White noise power spectrum, (b) Pink noise power spectrum.}
\label{fig:transforms}
\end{figure}

\begin{figure}[t!]
\centering
\includegraphics[width=.5\textwidth,height=.25\textheight]{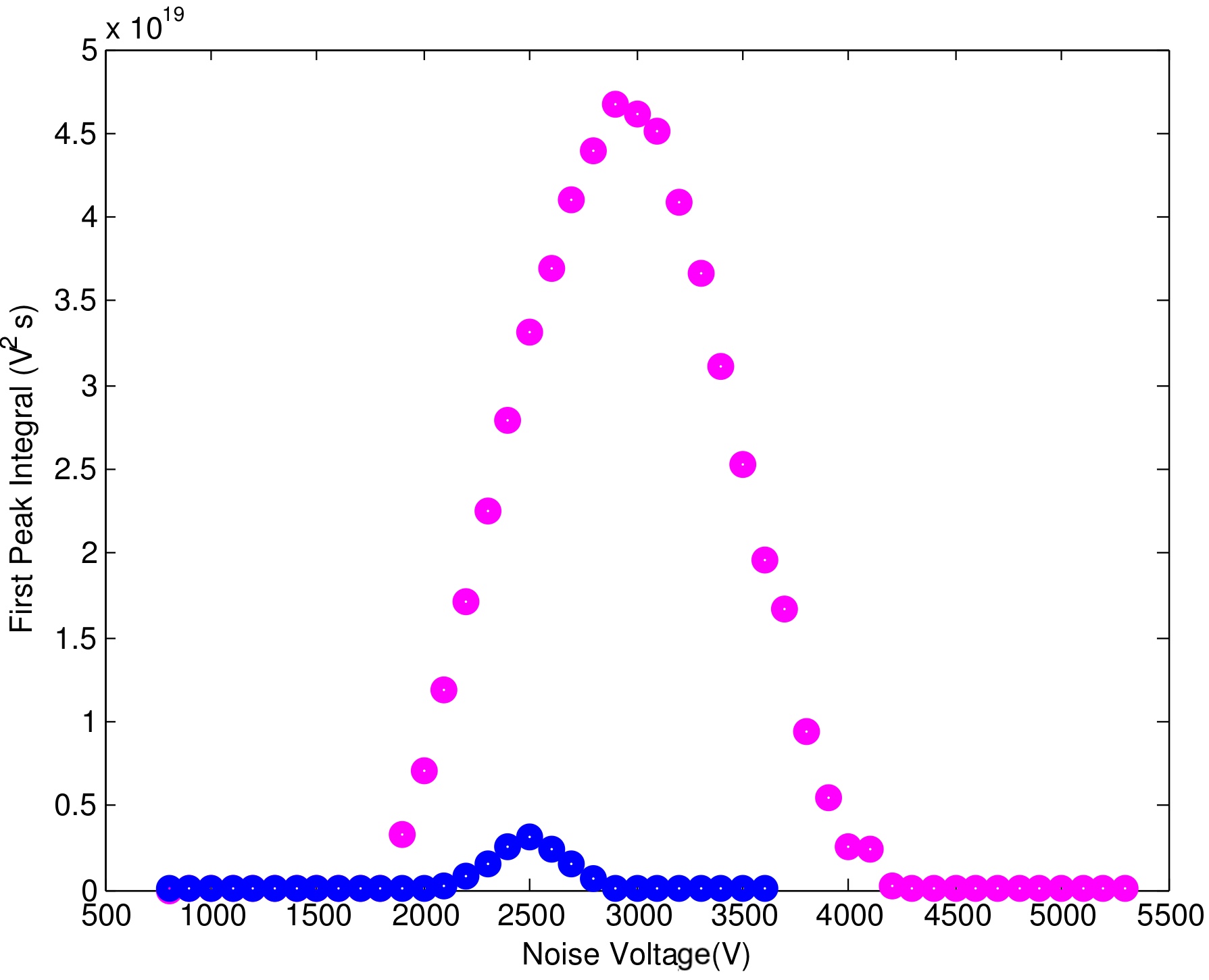}
\caption{Pink noise contribution to input signal frequency is 100 times smaller than white noise, however, pink noise amplifies the weak input signal 20 times more than white noise.}
\label{fig:comp}
\end{figure}

Fig. \ref{fig:transforms} (a) and (b) show the power spectrums for white and pink noises, respectively. It can be seen that the contribution of the input signal frequency corresponding to pink noise, is 100 times less than the spectrum of white noise. It is a surprising finding according to Fig. \ref{fig:comp}. Although the contribution of pink noise in the input signal frequency is 100 times less than white noise amplification of the weak input signal for pink noise is 20 times bigger than for white noise. This result is beneficial since numerous physical and biological phenomena are immerse in pink noise. Here, I poof that their amplification factor is high due to their nature and intrinsic characteristics.
It is important to notice that theoretical models, numerical and computational have used white noise since there is a mathematical algorithm to generate it, while there is no mathematical algorithm that is able to generate pink noise. According to this finding, the importance of pink noise to improve amplification factor is clear and it should be implemented in physical models to achieve higher accuracy. It is significant to mention that there is no bibliography which reported this relevant finding and here is the first report about it.

\section{Conclusion} \label{sec5}
In this paper, the SR model for an artificial neuron was implemented. Two schemes for the output signal analysis were used. The SNR analysis method which requires the integration of the maximum peaks is easier than the ISIH method since ISIH analysis, its results are not as precise as the SNR analysis scheme. Moreover, it is not recommended to use programming to obtain the Fourier transform of the output signal, since there are complications for programming and under slight parameter variations, which change the results rapidly. However, using spectrum analyser directly is faster and more reliable.\\ \\
It proved that the SR model amplifies the weak periodic input signal through noise. Thus, the SR model conserves the information of input signals. It shows that the signal is embedded in a noisy environment with the contents of all frequency values and with higher amplitudes than the desired signal. In the detection process, the frequency of the input signal and its multiples are retrieved. All measurements are replicated to validate the reliability of the results. It can be seen that the SR model can detect the input signal when the amplitude of this signal is decreased from 800 mv to 40 mv. Hence, it confirms that the SR model can conserve weak signals and enhance their amplitude. Moreover, the most important conclusion is that an artificial neuron can be modelled as a SR system in the presence of white or pink noise. Pink noise provides a higher performance in terms of amplification factor. Furthermore, pink noise has a wider range for optimum values, while white noise has a narrow optimum range. Hence, it is concluded that neurons are more sensitive to detect the signals which carry pink noise than signals with white noise or without noise. In addition to artificial neurons, the concluded results can be used in all types of systems with a threshold value and excitability behaviour where detection of weak signals embedded in a high amplitude noise is desirable. Since the sensory neuron's noise is limited by external and intrinsic neural noise, it is possible to add pink noise to the signals which are not detectable to retrieve by neurons. Thus, the neurons can detect weak signals, and thus, an extra-sensorial capability is created in a natural way.
\section{Acknowledgements}
I would like to give special acknowledgement to Keyvan Aghababaiyan, who motivated me to publish the results of this research and who helped me translating and edite it. Special gratitude to Professor Juan Gabriel Ramirez, who guided me and advised me during the experimental development of this research. Thank you so much Marco Gonzalez, for opening me the doors of his laboratory at unlimited times and without reservations and for teaching me data acquisition and analysis techniques. Special thanks Professor Rodolfo Llinas Riascos for inspiring me to study the Nervous System and for motivating me to pursued the unanswered questions. Special gratitude Professor Andres Reyes Lega, the main advisor of this research, his comments and unconditional support were the theoretical foundation of this project. This research was not funded by any institution.

\bibliography{Final}
%\nocite{*}

\end{document}